# Game Networks

**Pierfrancesco La Mura**
Dept. of Computer Science
Stanford University
Stanford, CA 94305, USA
plamura@stanford.edu

## Abstract

We introduce Game networks (G nets), a novel representation for multi-agent decision problems. Compared to other game-theoretic representations, such as strategic or extensive forms, G nets are more structured and more compact; more fundamentally, G nets constitute a computationally advantageous framework for strategic inference, as both probability and utility independencies are captured in the structure of the network and can be exploited in order to simplify the inference process. An important aspect of multi-agent reasoning is the identification of some or all of the strategic equilibria in a game; we present original convergence methods for strategic equilibrium which can take advantage of strategic separabilities in the G net structure in order to simplify the computations. Specifically, we describe a method which identifies a unique equilibrium as a function of the game payoffs, and one which identifies all equilibria.

## 1 Introduction

The formal analysis of multi-agent systems is a topic of interest to both economic theory and artificial intelligence. While game-theoretic notions and methodologies have already populated the economic mainstream, only recently they started to attract interest in the context of artificial intelligence, where their integration with existing methods constitutes a promising area of new research.

In this paper we introduce a new class of graphical representations for multi-agent decision problems, Game networks (G nets). Compared to standard game-theoretic representations, such as strategic and extensive forms, G nets are more structured and more compact, as both probabilities and utilities enjoy a modular representation. More fundamentally, G nets provide a computationally advantageous framework for strategic inference, as one can exploit conditional probability and utility independencies to reduce the complexity of the inference process.

An important aspect of multi-agent reasoning is the identification of some or all of the strategic equilibria in a game. For all but the simplest classes of games this a computationally demanding task, which can in principle be alleviated by making more efficient use of the information contained in the game structure (see [MM96] for a survey on the recent state of the art on the computation of strategic equilibria). We derive original convergence methods for strategic equilibrium which can exploit strategic separabilities in the G net representation in order to simplify computations. Specifically, we describe a path-tracking method which identifies a unique equilibrium as a function of the game payoffs, and one which identifies all equilibria.

G nets are closely related to a novel representation for single-agent decision problems, Expected Utility networks (EU nets), introduced in [LMS99]. EU nets generalize Markovian networks from the AI literature [Pea88], and provide a modular and compact framework for strategic inference.

EU nets exploit a novel notion of utility independence, closely related to its probabilistic counterpart. Together, conditional probability and utility independence imply conditional expected utility (or *strategic*) independence. What is important about strategically independent decisions is that they can be effectively decentralized: a single, complicated agent can be replaced by simpler, conditionally independent sub-agents, who can do just as well. This property is of interest not only to artificial intelligence, since it can be exploited to reduce the complexity of planning, but also to economic theory, as it suggests a principled way for the identification of optimal task allocations within economic organizations.

Yet, EU nets are somewhat limited for the purpose of modeling, as they do not capture causal relationships among events. Obviating to such limitation is a primary motivation for the introduction of G nets.

Game networks stand in roughly the same relationship to Bayesian networks as Expected Utility networks to Markovian networks. Bayesian networks encompass a probabilistic notion of causality, whereas the causal "parents" of each variable in the network are assumed to determine its conditional probability, but not its truth value. While the resulting implicit representation of the state space affords significant computational advantages, it still captures all the relevant information for the purpose of Bayesian inference, and provides an intuitive and compact modeling framework. Moreover, the probabilistic independencies captured by the network structure can be exploited to simplify the inference process, either directly or by first passing to a Markovian representation.

G nets encompass a decision-theoretic notion of causality, *Bayesian rationality*, whereas the agents' preferences constrain the conditional probabilities of the possible actions, but not their truth values. We argue that Bayesian rationality captures all the relevant information for the purpose of strategic inference, while at the same time avoiding some difficulties faced by other notions of causality. As in the case of Bayesian networks, one can either perform strategic inference directly on a G net, or switch to a (multi-agent) EU net representation first.

The rest of the paper is organized as follows. In section 2 we briefly present the decision-theoretic background of G nets. In section 3 we give a formal definition of G nets, and discuss how they can be used to model multi-agent decision problems in the context of a simple example. Finally, in section 4 we give existence and convergence results for strategic equilibrium in G nets.

## 2 Decision-theoretic preliminaries

In this section we present the decision-theoretic background of G nets; we first give a brief summary of the framework developed in [LMS98], and then discuss how to incorporate an appropriate notion of causality within the framework.

Let $\mathcal{A}$ be a Boolean algebra of events (or propositions), and $\succsim$ a preference ordering on the nonempty events $\mathcal{A} - \{\emptyset\}$. Without loss of generality, we extend $\succsim$ to the entire algebra $\mathcal{A}$ by assuming $E \succ \emptyset$ for all nonempty $E \in \mathcal{A}$.

We now define what it means for a preference ordering $\succsim$ to admit an expected utility representation.

**Definition 1** *An* expected utility representation *of* $\succsim$ *on* $\mathcal{A}$ *is a pair* $(P, U)$, *where* $P : \mathcal{A} \to [0, 1]$ *is a probability function and* $U : \mathcal{A} \to \mathbb{R}_+$ *is a non-negative utility function such that:*

1. $U(E) \geq U(F)$ *if and only if* $E \succsim F$

2. $U(E)P(E) = \sum_k U(E_k)P(E_k)$ *for any finite, measurable partition* $\{E_k\}$ *of* $E$.

Without loss of generality, we assume that the utility of the empty set is zero.

**Example 1** *Let $\Delta$ be the set of all probability measures defined on a finite algebra $\mathcal{A}$, and let $\succsim$ be a preference ordering on $\Delta$ which admits a von Neumann-Morgenstern expected utility representation. Let $P \in \Delta$ be a given, strictly positive probability measure representing the agent's prior beliefs, and for all nonempty $E \in \mathcal{A}$ let $P_E$ be the conditional measure defined by $P_E(F) = P(E \cap F)/P(E)$, $F \in \mathcal{A}$. Preferences over posterior probabilities $\{P_E\}$ can be identified with preferences over nonempty events in $\mathcal{A}$ by imposing $E \succsim F$ if and only if $P_E \succsim P_F$. Then the resulting preference ordering on $\mathcal{A}$ admits an expected utility representation in our sense.*

We say that a family $\{p_E\}_{E \in \mathcal{A} - \{\emptyset\}}$ is a *conditional probability system* if it assigns to every non-empty conditioning event $E \in \mathcal{A}$ a conditional probability measure over $\mathcal{A} \cap E$, such that the $p_E$ agree with Bayesian conditioning whenever possible; specifically, for any nonempty $E, F, G \in \mathcal{A}$ such that $G \subset F \subset E$, one has that $p_E(G) = p_E(F)p_F(G)$.

**Definition 2** *A* conditional expected utility representation *of* $\succsim$ *on* $\mathcal{A}$ *is a pair* $(p, u)$, *where* $p := \{p_E\}_{E \in \mathcal{A} - \{\emptyset\}}$ *is a conditional probability system and* $u : \mathcal{A} \to \mathbb{R}_+$ *is a non-negative utility function such that:*

- $u(E) \geq u(F)$ *if and only if* $E \succsim F$

- $u(E) = \sum u(E_k) p_E(E_k)$ *for any finite, measurable partition* $\{E_k\}$ *of* $E$.

Let $I$ be a finite set of agents. Agent $i \in I$ is assumed to have preferences not only about the basic events in a finite algebra $\mathcal{A}_0$ and its own preferences, but on other agents' preferences as well.

Let $W$ be a set of possible worlds, and $\succsim^i_w$ a function which associates to each pair $(i, w)$ a preference ordering on $2^W$. For any $E, F \in \mathcal{A}_0$, let $[E \succ^i F]$ denote the proposition $\{w \in W \mid E \succ_w^i F\}$ ("$i$ (strictly) prefers $E$ to $F$"), and by $[E \sim^i F]$ the set $\{w \in W \mid E \sim_w^i F\}$ ("$i$ is indifferent between $E$ and $F$"). Furthermore, let $\mathcal{B}_0^i$ denote the set of all finite intersections of propositions $[E \succ^i F]$ and $[E \sim^i F]$.

We recursively define $n$-th order ($n > 0$) algebras and preferences:

- $\mathcal{A}_n = \mathcal{A}_{n-1} \cup (\cup_{i \in I} \mathcal{B}_{n-1}^i)$ is the algebra generated by events in $\mathcal{A}_{n-1}$ and $\mathcal{B}_{n-1}^i$, $i \in I$

- $\mathcal{B}_n^i$ ($i \in I$) is the set of all finite intersections of propositions $[E \succ^i F]$, $[E \sim^i F]$, where $E, F \in \mathcal{A}_n$

- $\mathcal{A} = \cup_n \mathcal{A}_n$ is the algebra generated by events $E \in \mathcal{A}_n$ ($n \geq 0$), and $\mathcal{B}^i$ is the set of all finite intersections of propositions $[E \succ^i F]$, $[E \sim^i F]$, where $E, F \in \mathcal{A}$.

Notice that $\mathcal{B}^i \subset \mathcal{A}$, and hence $\mathcal{A} = \mathcal{A} \cup (\cup_{i \in I} \mathcal{B}^i)$. Therefore, further iteration is superfluous: all the preferences on events in $\mathcal{A}$ are already included in $\mathcal{A}$.

Under regularity conditions described in [LMS98] there exists an expected utility representation of agent $i$'s preferences on events in $\mathcal{A}_n$, for all finite $n$ and for all $i \in I$, which satisfies

$$U^i(E) P^i(E) = \sum U^i(E_k) P^i(E_k)$$

for any finite, $\mathcal{A}_n$-measurable partition $\{E_k\}$ of $E$. Observe that, since here probabilities and utilities emerge purely as expressions of preferences, statements about (unobservable, but strategically relevant) probabilities and utilities can be made sense of as statements about preferences (observable, at least in principle). Also, statements about higher-order probabilities and utilities (such as "I believe that you consider $E$ more likely than $F$", or "I would like you to believe $F$", etc., also useful when carrying out explicit strategic reasoning) can be similarly interpreted as statements about higher-order preferences.

Furthermore, under additional regularity conditions, for all $i \in I$ there exists a conditional expected utility representation of $\succsim$ on $\mathcal{A}$ which satisfies

$$u^i(E \cap F) = \sum u^i(E_k \cap F) p_E^i(E_k)$$

for all $(E, F) \in \mathcal{A} \times \mathcal{B}^i$, and for any finite, measurable partition $\{E_k\}$ of $E$.[1]

---
[1] For a precise statement of the result and for proofs see [LM99].

So far the representation does not involve any notion of causality: in other words, nothing links the agents' preferences on events with the actual occurrence of those events. To bring causality into the picture we first assume that the set of basic events $\mathcal{A}_0$ is the Boolean algebra generated by the available moves in an extensive form game. For each information set $H$ in the extensive form, let $\{E_H\}$ be the partition generated by the available moves at $H$. We shall refer to the $E_H$ as *optional events*, or *actions*.

We could now assume that, for any information set $H$ associated to agent $i$, and for any $E_H$ and $F_H$ which are available actions at $H$, $E_H \succ^i F_H$ implies $F_H = \emptyset$; in other words, we could regard any action which is not the most preferred by the agent at the corresponding information set as impossible. One problem with this approach is that all dominated actions, being impossible, must be indifferent to each other. In other words, for no three actions $E_H, F_H, G_H$ can it be the case that $E_H \succ^i F_H \succ^i G_H$. This naive treatment of causality leads to significant difficulties when reasoning about weakly dominated strategies, as the latter cannot be regarded as disbelieved yet possible events.

We avoid this problem by adopting a probabilistic notion of causality. In this view, all optional events are possible; the notion of possibility logically precedes the agents' decisions, and is part of the description of the game. In particular, this will allow us to regard a prisoner's dilemma, where cooperation is indeed an available action, as logically distinct from a degenerate game in which the two players have no choice at all.

The following property links preferences to actions; for any two actions $E_H, F_H$ available to agent $i$ at $H$, $E_H \succ^i F_H$ implies $p_H(F_H) = 0$. The interpretation of this assumption, which we shall refer to as *Bayesian rationality*, is that agents do not directly assign the truth values of optional events; rather, their preferences constrain the conditional probabilities of the possible actions in such a way that the probability of any dominated action is equal to zero. Intuitively, we are assuming that the agents' decisions may be prone to occasional mistakes, but the likelihood of any such mistake is infinitesimal.

## 3 Game Networks

There are two standard representations for mathematical games: the normal (or strategic) form, and the extensive form. The extensive form is more structured than the normal form: not only it describes the identities of the players, the strategies available to each player, and the payoff functions, as the normal form does; but also the information held by the agents at any possible state of the system, and the causal struc-ture of events in the game. The extensive form is at least as general as the normal form: any game in normal form can be interpreted as a game with simultaneous moves, and represented in the extensive form. The reverse operation (from the extensive to the normal form) is also possible, but some of the structural information is lost in the process, and hence the resulting normal-form representation is often impractically large.

Even though extensive-form representations are more compact than normal-form ones, they may still be quite redundant: often, in concrete examples, different end nodes may have identical payoffs, and the same actions may be available at different information sets, but the recognition of those symmetries does not lead to a more parsimonious representation. Furthermore, changing a few details in the setup usually entails rewriting the whole game; in other words, the extensive-form representation is not particularly modular. As we shall see, compared to extensive forms G nets are at least as general, but more structured and more compact.

In order to formally define G nets we need some more notation, which we borrow from [LMS99]. Let $\{X_k\}_{k \in N}$ be a set of variables, and for any $M \subset N$ let $X_M := \times_{k \in M} X_k$. Let $u : X_N \to \mathbb{R}_+$ be a strictly positive utility function, and let $\{A, B, C\}$ be a partition of $N$; also, define $-A := B \cup C$. Finally, let $w(x_A|x_{-A}) = u(x_A, x_{-A})/u(x_A^0, x_{-A})$, where $x^0 \in X_N$ is an arbitrary reference point; $w(x_A|x_{-A})$ is called the *utility potential* at $X_A$, and captures how the utility changes when the variables in $A$ shift away from the reference point, while the values of the remaning variables are held fixed at $x_{-A}$. If, for all $(x_A, x_{-A})$, the value of $w(x_A|x_{-A})$ only depends on $x_C$, and not on $x_B$, we say that $X_A$ is $u$-independent of $X_B$ given $X_C$, and define new quantities $w(x_A|x_C) := w(x_A|x_B^0, x_C)$. If $X_A$ and $X_B$ are both probabilistically and $u$-independent given $X_C$ then we say that they are *strategically independent* given $X_C$.

### 3.1 A formal definition

G nets are comprised of a finite, ordered set of nodes $X_N$ (where $N = \{1, ..., n\}$), corresponding to a set of strategically relevant variables which we also assume to be finitely-valued, a partition $I$ of $N$ which determines the identity of the agent responsible for the decision at each node[2] (including Nature), and two types of arc, representing causal and preferential (teleological) de-pendencies. Causal dependencies are represented by directed (probability) arcs, with no cycles, and preferential dependencies by undirected (utility) arcs.

---

[2]For simplicity, we assume that all the actions available at node $k$ pertain to the same player $i(k)$, although this assumption can be relaxed. In fact, our treatment carries over without changes if we assume that different agents are active at different information sets of the same node.

A node $X_k$ ($k \in N$) is associated with two quantities, $w(x_k|x_{UN(k)})$ and $p(x_k|x_{PF(k)})$, where $UN(k)$ are the nodes directly connected to $X_k$ via utility arcs (the utility neighbors of $X_k$), and $PF(k)$ is the set of probability parents [3] of $X_k$. While the $p(x_k|x_{PF(k)})$ identify a conditional probability system, the same for all players, $w$ is a vector of functions $(w^i)_{i \in I}$, one for each player. In turn, the $w_i(x_k|x_{UN(k)})$ are interpreted as the utility potentials coming from some strictly positive utility function $u^i$.

The incoming probability arcs represent those events which the agent who controls $X_k$ can observe at the moment of decision. A decision-maker (including Nature) may choose any random rule, as long as it depends on the truth values of the $X_{PF(k)}$ only. An element of the partition generated by the $X_{PF(k)}$ is called an *information set* at $k$, and represents all the information available to the agent at the moment of decision.

We say that payoffs are *normal* if the utilities of all states are positive, and are expressed as multiples of $u^i(x^0)$ (where $x^0$ is the arbitrary reference point). Clearly, any game can be transformed in one with normal payoffs via a suitable positive affine rescaling of the original payoffs. Hence, without loss of generality, we shall concentrate on games with normal payoffs.

A G net in which only the probabilities of Nature's actions (and not those of the other agents') are specified is a *G frame*. A G frame can be regarded as the set of all G nets which respect the implied independence structure, and agree in the utility assignments and in the probabilities of Nature's actions.

**Theorem 1** *Any finite game in extensive form has a G frame representation.*

Let $A_i(H)$ be the set of actions available to player $i$ at an information set $H$. $A_i(H)$ can be regarded as a partition of $H$ into possible actions $E_H$.

**Definition 3** *A Game network is said to satisfy Bayesian rationality if, for all $k \in N$, for all information sets $H = x_{PF(k)}$, and for any $E_H, F_H \in A_{i(k)}(H)$, it is the case that $u^{i(k)}(E_H|H) > u^{i(k)}(F_H|H)$ implies $p_H(F_H) = 0$.*

Anticipating our discussion on the existence of strategic equilibrium in game networks, we state the follow-ing important corollary.

---

[3]The probability parents of a node $X_k$ are those nodes which are immediate predecessors of $X_k$ in the partial ordering induced by the (directed) probability arcs.

**Corollary 2** *For any finite G frame there exists a corresponding G net which satisfies Bayesian rationality.*

### 3.2 An example: the beer/quiche game

As an example of G net we present the beer/quiche game. In this game, Nature selects the type of player 1, who may be either strong (S) or weak (-S). Player 1 (who knows his type) goes to a pub, and has to decide whether to get beer (B) or quiche (-B). His decisions are observed by player 2, who is a bully and, as such, enjoys fighting against weak types. After observing what player 1 orders, player 2 decides whether to start a fight with player 1 (F) or not (-F). If player 1 is strong he will fight back, and hence player 2 in that case prefers not to fight. Player 1 always prefers not to get into a fight, but more strongly so if he is a weak type and hence knows that he would be beaten up. Finally, strong types of player 1 prefer beer to quiche, while weak types have the opposite preference.

A G net (or, more precisely, a G frame) representation of this game is depicted in figure 1, along with the corresponding extensive-form representation. The probability dependencies are represented by solid arrows, while the dashed and dotted lines represent utility dependencies for players 1 and 2 respectively.

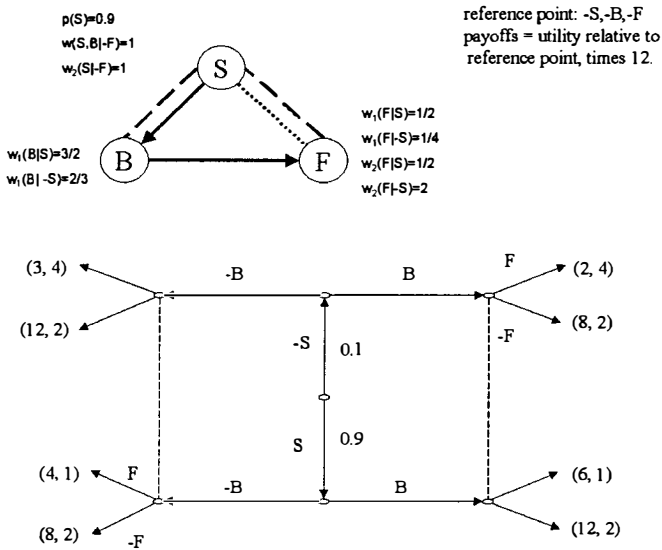

Figure 1: The beer/quiche game.

In the informal description above are buried several independence assumptions. For instance, it is implicitly assumed that Nature's choice cannot depend on what the two players will do later, or that player 2's decision contingent on the observation of player 1's behavior is independent of Nature's choice. Moreover, it is implicitly assumed that player 1 prefers beer or quiche regardless of whether he will have to fight or not, or that player 2 only cares about 1's type if he chooses to fight, but not otherwise.

In the G net representation, both probability and utility independencies are captured in the structure of the network. By contrast, the extensive form only captures probability independencies, while the recognition of utility independencies (which induce symmetries in the payoff structure) does not lead to a more compact representation. In our example, compactness is also reflected in the number of parameters needed to specify the game payoffs: in the extensive form one needs 16 parameters to identify the payoffs, while in the G net representation one only needs 8.

Compactness is only one of several advantages of G nets over extensive forms; another one is modularity. For instance, in a G net one can easily introduce new moves (e.g., the reaction of a third player to player 2's decision to fight or not), or change the informational assumptions while at the same time retaining most of the existing structure (in particular, payoffs do not need to be completely reassessed if the state space is refined).

A third advantageous feature of G nets is that the relevant information about utilities can generally be introduced more naturally than in extensive forms. In the context of our example, for instance, going from the informal description to a numeric assessment of the payoffs is relatively cumbersome; the decision maker needs to report absolute utility values for all possible outcomes, while the informal description only compares a few different scenarios. By contrast, to construct a G net one only needs information about payoff dependencies and order-of-magnitude comparisons between the relative utilities of alternative scenarios, which is closely related to the type of information contained in the informal description.

## 4 Strategic equilibrium in game networks

An important aspect of multi-agent reasoning is the identification of some or all of the strategic equilibria in a game. In this section we establish convergence results for strategic equilibrium which can exploit strategic separabilities in the G net representation in order to simplify the computation of equilibria. Specifically, we describe a path-tracking method which identifies a unique equilibrium as a function of the game payoffs, and one which identifies all equilibria.

Let $X$ be a finite set of states, and let $\mathcal{A}$ be a Boolean algebra of subsets of $X$.

We assume that the agent's preferences on $\mathcal{A}$ admit a conditional expected utility representation $(p, u)$, where $p$ is a conditional probability system defined on $\mathcal{A}$, and $u$ is a function associating to each state $x$ a positive real number $u(x)$.

The expected utility and conditional expected utility for general events are defined by

$$u(p)(E) := \sum_x u(x)p(x|E), \text{ and}$$

$$u(p)(F|E) := \frac{u(E \cap F)}{u(E)}.$$

Let $v(F|E) := p(F|E)u(F|E)$; we say that $v_E(F) := v(F|E)$ is the conditional *value* of $F$ given $E$. Notice that $\{v_E\}_{E \in \mathcal{A}-\{\emptyset\}}$ is a conditional probability system. Moreover, for any nonempty conditioning event $E$ and for any $F$ whose conditional probability is nonzero,

$$u(F|E) = \frac{v(F|E)}{p(F|E)}.$$

Hence, in such cases we can regard the conditional expected utility as the ratio of two conditional probabilities, one representing value and the other belief. Note that, even when the probability of $F$ given $E$ is equal to zero, the conditional expected utility $u(p)(E \cap F)/u(p)(E)$ is well defined and strictly positive.

### 4.1 Existence of equilibrium in game networks

Let $G$ be a game network characterized by a finite set $X = \{X_k\}_{k \in N}$ of decision nodes and their associated probability (directed) and utility (undirected) arcs, and let $u^i : X \to \mathbb{R}_+$ be the utility function for player $i$ ($i \in I$). For each $k \in N$, let $i(k)$ denote the player responsible for the choice of $X_k$. Let $\Delta$ be the set of all conditional probability systems $\{p_H\}$ on $X$ where $H$ is an information set $x_{PP(k)}$ for some $k \in N$, and $PP(k)$ are the probability parents of $k$. Then the probability of any state $x$ can be obtained by means of the simple product rule $p(x) = \times_{k \in N} p(x_k|x_{PP(k)})$. Notice that $\Delta$ is a compact subset of $\mathbb{R}^n$ (where $n$ is the number of parameters $p(x_k|x_{PP(k)})$ which characterize $p$), and is also convex with respect to combinations $p\lambda p'$ defined by

$$p\lambda p'(x_k|x_{PP(k)}) = \lambda p(x_k|x_{PP(k)}) + (1-\lambda)p'(x_k|x_{PP(k)}),$$

for $k \in N$, $\lambda \in [0, 1]$.

Next, let $v : \Delta \to \Delta$ be a function which associates to each $p \in \Delta$ a conditional probability system $\{v_H\}$, where $H = x_{PP(k)}$ are the information sets, and

$$v(p)(x_k|x_{PP(k)}) = p(x_k|x_{PP(k)})u^{i(k)}(p)(x_k|x_{PP(k)}).$$

Then $v$ is a continuous self-function on a convex and compact subset of $\mathbb{R}^n$, and hence it has a fixed point by Brouwer's theorem. A fixed point is characterized by the set of equalities $p(x_k|x_{PP(k)}) = v(p)(x_k|x_{PP(k)})$, for all $k$, $x_k$ and $x_{PP(k)}$. Let $F$ be the set of such fixed points.

A *Nash equilibrium* is defined as a conditional probability system $p \in \Delta$ such that

$$\sum_{x_k} p(x_k|x_{PP(k)})u^{i(k)}(p)(x_k, x_{PP(k)})$$
$$\geq \sum_{x_k} q(x_k|x_{PP(k)})u^{i(k)}(p)(x_k, x_{PP(k)})$$

for all $k \in N$, $x_{PP(k)}$ and $q \in \Delta$. The set of Nash equilibria $E$ is contained in $F$, as equilibrium probabilities satisfy the fixed point conditions $p(x_k|x_{PP(k)}) = v(p)(x_k|x_{PP(k)})$. This is an immediate consequence of the following result.

**Proposition 3** *$p$ is a Nash equilibrium if and only if*

1. $u^{i(k)}(p)(x_k|x_{PP(k)}) \leq 1$, and

2. $u^{i(k)}(p)(x_k|x_{PP(k)}) = 1$ if $p(x_k|x_{PP(k)}) > 0$.

How do we know that the set of Nash equilibria is nonempty? It is easy to check that Nash equilibria in G nets correspond to Nash equilibria in the agent-strategic form [FT91, p.354], and hence an equilibrium exists by Nash's theorem. Yet, we present a simple direct proof, which motivates the convergence method we define later on.

Let $f_\varepsilon : \Delta \to \Delta$ be the function defined by $f := z \varepsilon v = \varepsilon z + (1-\varepsilon)v$, where $z$ is the conditional probability system which assigns equal probability to all the available actions at each information set. Brouwer's theorem guarantees that the set of fixed points of $f_\varepsilon$ is not empty. Fixed points of $f_\varepsilon$ have an important property:

**Proposition 4** *If $p$ is a fixed point of $f_\varepsilon$, then $u^{i(k)}(p)(x_k|x_{PP(k)}) > u^{i(k)}(p)(x'_k|x_{PP(k)})$ implies $p(x_k|x_{PP(k)}) > p(x'_k|x_{PP(k)})$.*

We define a *robust equilibrium* as a limit point of a sequence of fixed points of $f_\varepsilon$, as $\varepsilon$ goes to zero. By compactness of $\Delta$ any such sequence has a limit point, and hence the set of robust equilibria is nonempty. Since a robust equilibrium always exists, the following result ensures that the set of Nash equilibria is not empty.

**Proposition 5** *Any robust equilibrium is a Nash equilibrium.*

In fact, a stronger result holds: robust equilibria are also perfect, as they are easily shown to satisfy Definition 8.5C in [FT91, p.352].

Robust equilibria are similar to proper equilibria [Mye78], in that they are limits of sequences of strictly positive measures such that strategies with higher utilities always have higher probabilities.

### 4.2 Global convergence to equilibrium in game networks

Let $\{u^i\}_{i \in I}$ be the utility functions in a G net. We saw that an equilibrium is a conditional probability system $p$ which satisfies $F(p) = 0$, where $F(p)$ is the vector of functions defined by

$$F(p)(x_k|x_{PP(k)}) = p(x_k|x_{PP(k)}) - v(p)(x_k|x_{PP(k)}), \text{ and}$$

$$v(p)(x_k|x_{PP(k)}) = p(x_k|x_{PP(k)}) \frac{u^{i(k)}(p)(x_k, x_{PP(k)})}{u^{i(k)}(p)(x_{PP(k)})}.$$

We study convergence to a zero of the vector $F$ under the assumption that the probability of an action increases or decreases in proportion to its relative utility with respect to the other available actions. For now we shall ignore the fact that some fixed points may fail to be equilibria; in fact, we show below that the method we are presenting will converge to a Nash equilibrium in games with generic payoffs.

Consider the perturbed problem

$$F_\varepsilon(p)(x_k|x_{PP(k)}) = p(x_k|x_{PP(k)}) - f_\varepsilon(p)(x_k|x_{PP(k)}).$$

Observe that $F_\varepsilon$ can be rewritten as $\varepsilon F^0(p) + (1-\varepsilon)F(p)$, where $F(p)$ is the target system whose zeros we want to find and $F^0(p)$ is the trivial system $(p(x_k|x_{PP(k)}) - z(x_k|x_{PP(k)}))_{x_k, x_{PP(k)}}$, whose unique solution is $p = z$. Then $F_\varepsilon$ defines a convex-linear homotopy $h(p,t) = F_{1-t}$ [Mor87, p.135] with parameter $t \in [0,1]$. Note that $h$ coincides with the trivial system for $t = 0$, and with the target system for $t = 1$.

In our setting $h$ is extremely well behaved: for generic payoffs it satisfies conditions 1,2,3 and 4b in [Mor87, p.122] by construction, and moreover it satisfies condition 5 (in $\mathbb{R}^n$) because $F$ is bounded. Therefore, standard path-tracking methods are guaranteed to work for generic games. The end point of the homotopy path is a robust equilibrium, as it is the limit, as $\varepsilon$ goes to zero, of a sequence of solutions for the perturbed problem. To handle degenerate cases, in which the uniform distribution is a bad choice of initial condition, it suffices to introduce a slight random perturbation to the game payoffs to guarantee convergence.

Now we can define a new solution concept (the end point of the homotopy path), which we name *first equilibrium*, and claim that:

- a generic G net has a unique first equilibrium
- the first equilibrium is uniquely determined by the payoff structure of the game
- the first equilibrium of a generic G net can be approximated using standard path-tracking methods
- the first equilibrium is a robust equilibrium of the game.

How does the above procedure compare with the existing game-theoretic methods for computing a sample equilibrium? The literature on computational methods in game theory is quite technical, and we shall not attempt an explicit comparison here; for a survey on the recent state of the art we refer to [MM96]. The main advantage of our approach with respect to other methods is that, for given $(x_k, x_{PP(k)})$, strategically independent variables do not affect the values of $F(p)(x_k|x_{PP(k)})$; it follows that, in the presence of strategic independencies, convergence to the zeros of a large system can be reduced to convergence to the zeros of smaller, strategically independent subsystems. The following example illustrates the point for a simple case.

**Example 2** *Suppose that there is only one agent, and let $A$ and $B$ be two strategically independent subsets of variables with marginal probability functions $p_a$ and $p_b$ respectively. Then $p = p_a \times p_b$, and $F(p)(a)$ and can be written as $p_a \left(1 - \frac{w(a|b_0)}{\sum_{a'} w(a'|b_0) p_{a'}}\right)$, while a similar expression holds for $F(p)(b)$. Note that $F(p)(a)$ (resp., $F(p)(b)$) is a function of $p_a$ (resp., $p_b$) only, and hence the zeros of $F(p)$ correspond to the zeros of the two independent subsystems $F_A(p_a) := F(p)(a)$ and $F_B(p_b) := F(p)(b)$.*

### 4.3 Computing all the equilibria

The convergence method we presented above only tracks a single robust equilibrium. Yet, in many cases one wants a complete list of all the Nash equilibria. For instance, we may want to design agents who always coordinate on a Pareto efficient equilibrium, whenever such efficient equilibrium is unique. How would our agents know if that condition is met? In principle, they will need to identify all the equilibria in order to infer their optimal strategies.

Finding all equilibria is computationally very demanding. As in the single-equilibrium case, we seek a method which can take advantage of strategic independencies in order to simplify computations. It turns out that an adaptation of the procedure defined in section 4.2 can be put to such use.

Let $G(p)$ be the vector of functions $(g_j)_{j \in J}$ (where $j = (x_k, x_{PF(k)})$) defined by $G(p)(x_k|x_{PF(k)}) = u^{i(k)}(p)(x_{PF(k)})F(p)(x_k|x_{PF(k)})$. Since $u$ is strictly positive and bounded, the zeros of $G$ coincide with the zeros of $F$.

Notice that both the numerator and the denominator of $v(p)(x_k|x_{PF(k)})$ are polynomial functions of the $p(x_k|x_{PF(k)})$, and hence $G(p)$ is a vector of polynomial functions, whose zeros include all the Nash equilibria. Also, observe that strategically independent variables do not affect the zeros of $G(p)(x_k|x_{PF(k)})$; they only play the role of multiplicative scaling factors. As in the single-equilibrium case, in the presence of strategic independencies we can replace a large system of polynomial equations with smaller, strategically independent subsystems, as the following example shows for a simple case.

**Example 3** *As in example 2 assume that there is a single agent, and two strategically independent subsets of variables $A$ and $B$. Then $G(p)(a)$ can be written as $p_a \left[\sum_{a'} w(a'|b_0)p_{a'} - w(a|b_0)\right] \left(\sum_b w(b|a_0)p_b\right)$, while a similar expression holds for $G(p)(b)$. Observe that the last term in the product is strictly positive, and hence the zeros of $G(p_a, p_b)$ correspond to the zeros of the two independent subsystems $G_A(p_a)$ and $G_B(p_b)$, where $G_A(p_a)(a) := p_a \left[\sum_{a'} w(a'|b_0)p_{a'} - w(a|b_0)\right]$, and $G_B(p_b)(b) := p_b \left[\sum_{b'} w(b'|a_0)p_{b'} - w(b|a_0)\right]$.*

A Nash equilibrium may not have any homotopy path converging to it in $\mathbb{R}^n$. Yet, the following result ensures that we can get at them in the complex space $\mathbb{C}^n$. Let $G^0$ be the initial system defined by

$$G_j^0(p) = \alpha_j^{d_j} p_j^{d_j} - \beta_j^{d_j},$$

where $j = (x_k, x_{PF(k)})$, $d_j$ is the degree of $G(p)(x_k|x_{PF(k)})$, and $\alpha_j$ and $\beta_j$ are generic complex constants. Then $G^0(p) = 0$ has $d = \times_{j \in J} d_j$ solutions. Let $h(x,t)$ be the homotopy defined by

$$h_j(p,t) = (1-t)G_j^0(p) + tG_j(p).$$

Then the following result in [Mor87, p. 60] applies.

**Theorem 6** *Given $G$, there are sets of measure zero, $A_\alpha$ and $A_\beta$ in $\mathbb{C}^n$ such that, if $\alpha \notin A_\alpha$ and $\beta \notin A_\beta$, then:*

1. *the solution set $\{(p,t) \in \mathbb{C}^n \times [0,1) : h(p,t) = 0\}$ is a collection of $d$ non-overlapping (smooth) paths*

2. *the paths move from $t = 0$ to $t = 1$ without backtracking in $t$*

3. *each geometrically isolated solution of $G = 0$ of multiplicity $m$ has exactly $m$ continuation paths converging to it.*

Theorem 6 guarantees that the homotopy paths generated by $h$ are well-behaved, and can be tracked with standard computational techniques. Observe that this method will identify all the zeros of $G$, including those which are not Nash equilibria; yet, the latter are easily singled out, as they correspond to the real solutions which lie in the region delimited by $u^{i(k)}(p)(x_k|x_{PF(k)}) \leq 1$.